\documentclass{article}
\usepackage{amsmath, amsthm, amssymb,  mathrsfs}
\usepackage{graphicx}
\usepackage{float}
\title{Optical 2-metrics of Schwarzschild-Tangherlini Spacetimes and the Bohlin-Arnold Duality}
\author{ }

\begin{document}
\maketitle

\begin{abstract}
In what follows, we consider the projection of null geodesics of the Schwarzschild-Tangherlini metric in $n+1$ dimensions to the space of orbits of the static Killing vector where the motion of a given light ray is seen to lie in a plane. The projected curves coincide with the unparametrised geodesics of optical 2-metrics and can be equally understood as describing the motion of a non-relativistic particle in a central force. We consider a duality between the projected null curves for pairs of values of $n$ and interpret its mathematical meaning in terms of the optical 2-metrics. The metrics are not projectively equivalent but the correspondence can be exposed in terms of a third order differential equation. We also explore the extension of this notion of duality to the Reissner-Nordstrom case.
\end{abstract}

\section{Introduction}
The study of the dynamics of null geodesics in black hole Lorentzian spacetimes has recently been approached in several contexts via the idea of projecting the curves to a spatial surface of lower dimension. For example, if a given metric admits a hypersurface-orthogonal timelike Killing vector $K$, then the null geodesics project down to the unparametrised geodesics of the \emph{optical metric} on the space of orbits of $K$. Similar constructions have been explored when the metric admits a stationary Killing vector \cite{stationary} or a timelike conformal retraction \cite{myself} where the projected null curves endow a hypersurface with some notion of geometric structure. In \cite{warnick}, the authors showed that the optical metrics of Schwarzschild-deSitter are \emph{projectively equivalent} (yield the same geodesics as unparametrised curves) for different values of the cosmological constant $\Lambda$ and, as a consequence, equations governing the dynamics of light rays in these spacetimes are independent of $\Lambda$. Thus, projective equivalence of optical metrics can play an important role in general relativity (see \cite{self} for a more detailed study of this topic). 
\\In \cite{vyska}, the authors consider the properties of null geodesics in Schwarzschild-Tangherlini spacetimes of $n+1$ dimensions. Here, the projection of any such curve to the space of orbits of the timelike Killing vector lies in a plane and coincides with an unparametrised geodesic of a two-dimensional optical metric. It is seen that the cases $n=3$ and $n=6$ may be related by a conformal mapping due to Bohlin \cite{bohlin} and Arnold \cite{arnold}. This begs the question as to whether the optical 2-metrics in these cases are projectively related and, if not, how can this relationship described?
\\In this paper, we explore the Bohlin-Arnold duality in depth in this context and argue that it does not give rise to projective equivalence of metrics but is something more general which can be described by a third order differential equation. We analyse the role of the cosmological constant for these spacetimes and show explicitly why it does not effect the equations governing the dynamics of light rays and we discuss how the zero energy solutions fit in to the duality. In Section 6, we consider the possibility of a similar notion of duality for Reissner-Nordstrom spacetimes in $n+1$ dimensions.

\renewcommand{\abstractname}{Acknowledgements}
\begin{abstract}
I would like to thank Prof. Gary Gibbons for introducing me to this problem and for his valuable input throughout.
\end{abstract}

\section{Null geodesics and Optical Metrics}
Ricci-flat black holes in $n+1$ dimensions can be described by the Schwarzschild-Tangherlini (ST) metric \cite{tangherlini}
\begin{equation}
g_{ST} = - \Delta dt^2 + \frac{dr^2}{\Delta} + r^2 d \Omega_{n-1}^2,
\label{first}
\end{equation}
where $d\Omega_{n}^2$ is the round metric on the unit $(n-1)$-sphere and
\begin{equation*}
\Delta = 1 -  \frac{2M_n}{r^{n-2}}.
\end{equation*}
If we project a null geodesic of this metric to the space of orbits of the Killing vector $\frac{\partial}{\partial t}$, we find that it lies entirely in a plane through the origin. Endowing this plane with polar coordinates $(r,\phi)$ and setting $u = \frac{1}{r}$ this motion is described by the differential equation
\begin{equation}
\left( \frac{du}{d\phi} \right)^2 + u^2 = 2 M_n u^n + \frac{1}{b^2}
\label{nullgeod}
\end{equation}
where $b$ is a constant impact parameter. Null geodesics of $g_{ST}$ may then be mapped into the motion of a non-relativistic particle moving in an attractive central force
\begin{equation*}
F \propto \frac{1}{r^{n+1}} \Leftrightarrow V \propto \frac{1}{r^n}.
\end{equation*}
Thus, one may use results from dynamics to discuss the optics of black holes (see \cite{belbruno} for a recent application of this idea). Alternatively, equation (\ref{nullgeod}) describes unparametrised geodesics of the optical 2-metric:
\begin{equation}
ds_{o_n}^2 = \frac{dr^2}{\Delta^2} + \frac{r^2}{\Delta} d\phi^2
\label{mainmet}
\end{equation}
with $0 \leq \phi \leq 2 \pi $, and projected null geodesics of the metric $g_{ST}$ precisely coincide with the totality of unparametrised geodesics described by (\ref{mainmet}) on each plane through the origin.
\\\textbf{Remark: } According to \cite{whittaker} and \cite{pars}, the cases $n = 3,4,6$ are integrable and may be solved in terms of elliptic functions. For a recent discussion, see \cite{hackmann}. The case $n=3$ admits a special solution in the form of a \emph{cardioid}. The case $n=6$ admits a special solution of the form of a \emph{Lemniscate of Bernoulli} with node at the singularity and which touches the horizon.

\section{Bohlin-Arnold Duality}
The following equivalence between dynamical systems in the plane is due to Arnold \cite{arnold} but it has its origin in a paper due to Bohlin \cite{bohlin}. He introduces the complex coordinate $z = x + iy$ and uses the Jacobi Principle, according to which, at fixed energy per unit mass $\mathcal{E}$, the paths described by (\ref{nullgeod}) will be unparametrised geodesics of the metric
\begin{equation}
ds_{\text{Jacobi}}^2 = \left( 2 \mathcal{E} - V(x,y) \right) dz d\bar{z}.
\label{jacobi}
\end{equation}
Now consider a similar system in the complex $w = u + iv$ plane with Jacobi metric
\begin{equation*}
ds_{\text{Jacobi}}^2 = \left( 2 \tilde{\mathcal{E}} - \tilde{V}(x,y) \right) dw d\bar{w}.
\end{equation*}
The two systems will coincide under pullback by the conformal map
\begin{equation*}
w = f(z)
\end{equation*}
if
\begin{equation*}
V = - |f'(z)|^2 \tilde{\mathcal{E}} \,\,\,,\,\,\, \mathcal{E} = -|f'(w)|^2 \tilde{V}.
\end{equation*}
Let us consider only conformal maps of the form $w = z^p$ for now. For these maps, one finds that $V \propto r^{2p-2}$ and $\tilde{V} \propto r^{\frac{2-2p}{p}}$ will work (setting $p \rightarrow \frac{1}{p}$ merely interchanges the role of $V$ and $\tilde{V}$). Furthermore, such expressions for the potential are physically interesting from the perspective of the classical orbit. $p=1$ gives a trivial case, but some other cases are of special note
\begin{itemize}
\item $p = -1$, i.e, \emph{inversion} which is the self-dual case with
\begin{equation*}
V \propto \tilde{V} \propto \frac{1}{r^4}.
\end{equation*}

\item $p=2$ takes the simple harmonic oscillator to the Kepler problem
    \begin{equation*}
    V \propto r^2 \,\,\,,\,\,\, \tilde{V} \propto \frac{1}{r}.
    \end{equation*}

\item $p = -\frac{1}{2}$, this example will be at the core of our discussion and gives
    \begin{equation*}
    V \propto \frac{1}{r^3} \,\,\,,\,\,\, \tilde{V} \propto \frac{1}{r^6}.
    \end{equation*}

\end{itemize}
Another useful way to look at things is to note that if
\begin{equation*}
\left( \frac{du}{d\phi} \right)^2 + u^2 = A u^{\alpha} + B
\end{equation*}
then
\begin{equation*}
\left( \frac{dr}{d\phi} \right)^2 + r^2 = A r^{4- \alpha} + B r^4.
\end{equation*}
This seems as though it would be particularly relevant for $\alpha = 4$. We will come back to the interpretation of this formulation later.
\\The duality between black holes in 3+1 and 6+1 dimensions is interesting. One example of this is the two pairs of special solutions
\begin{equation*}
au = \frac{\cosh{\phi} - 2}{\cosh{\phi} + 1} \,\, \text{or} \,\, \frac{\cosh{\phi} + 2}{\cosh{\phi} - 1} \,\,\, \Rightarrow F \propto \frac{1}{r^4}.
\end{equation*}
\begin{equation*}
a^2 u^2 = \frac{\cosh{2\phi} - 1}{\cosh{2 \phi} +2} \,\, \text{or} \,\, \frac{\cosh{2\phi} + 1}{\cosh{2 \phi} - 2} \,\,\, \Rightarrow F \propto \frac{1}{r^7}.
\end{equation*}
The first goes to the second under the replacement
\begin{equation*}
(au, \phi) \rightarrow (a^2 u^2, 2 \phi),
\end{equation*}
which is precisely Bohlin-Arnold duality.
\\The special zero energy solutions are not related in the same way. However, we will see why this is the case in the following sections.

\section{Projective Equivalence}
Now that we have demonstrated the Bohlin-Arnold duality, we may probe it a bit further. The results of the previous section seem to hint at the notion of projective equivalence. In particular, it appears that the metrics (\ref{mainmet}) for $n=3$ and $n=6$ may give rise to the same geodesics as unparametrised curves. Here, we present an argument for why that is not the case and explore the true consequences of the duality.
\\The family of metrics projectively equivalent to (\ref{mainmet}) can be completely determined for arbitrary $n \geq 3$ \footnote{We have verified this using the results in \cite{bryant}.} (obviously for $n=2$, the metric (\ref{mainmet}) is flat):
\begin{equation}
g_{n} = \frac{dr^2}{a \left( a-\frac{2aM_n}{r^{n-2}}+cr^2 \right)^2} + \frac{r^2}{a^2 \left( a-\frac{2aM_n}{r^{n-2}}+cr^2 \right)} d\phi^2
\label{projopt}
\end{equation}
for constants $a$ and $c$ (i.e, degree of mobility 2).
\\This more general metric (\ref{projopt}) is also familiar from a physical point of view. It is the optical metric of an ST black hole with cosmological constant $\Lambda = - \frac{n(n-1)}{2} c$. As mentioned in the introduction, the projective equivalence of such metrics was noted in \cite{warnick} where the invariance of the dynamics of light rays was discussed. In fact, we will see later that there is a link between projective invariance and the introduction of a cosmological constant in a more general case.
\\It is clear from (\ref{projopt}) that, in the given set of coordinates, the metrics for $n=3$ and $n=6$ are not projectively equivalent but, of course, may be so after a coordinate transformation.  In general, determining if such a diffeomorphism exists is a difficult problem. However, we can consider the case where just the $r$-coordinate is transformed - this is the type of transformation suggested by the Bohlin-Arnold duality. If we consider such a coordinate change for the $n=3$ metric (\ref{projopt}) which results in the $n=6$ case, given by $r = F(\tilde{r})$, then we find that it is impossible to transform both the $dr^2$ and $d\phi^2$ terms simultaneously in the appropriate way.
\\Hence, it seems that these metrics are not projectively equivalent even after a diffeomorphism.

\section{Probing the Duality}
The equation for unparametrised geodesics of (\ref{mainmet}) is
\begin{equation*}
r'' - \frac{2(r')^2}{r} + \frac{nM_n}{r^{n-3}} - r = 0
\end{equation*}
where ' represents differentiation with respect to $\phi$. Alternatively, we can express everything in terms of $u = \frac{1}{r}$:
\begin{equation}
u'' + u = nM_n u^{n-1}
\label{geodeq}
\end{equation}
or integrating once
\begin{equation}
(u')^2 + u^2 = 2M_n u^n + \frac{1}{b^2},
\label{impact}
\end{equation}
where $b$ is a constant impact parameter, as before. In this form, we can expose the correspondence between the $n=3$ and $n=6$ cases. Specifically, let $n=3$ in (\ref{impact}) and make the transformation
\begin{equation}
u = \tilde{u}^p = \tilde{u}^{-2} \,\,\,\,\,,\,\,\,\,\, \phi = p \tilde{\phi} = -2 \tilde{\phi}.
\label{transfo}
\end{equation}
Then (\ref{impact}) becomes
\begin{equation*}
(\tilde{u}')^2 + \tilde{u}^2 = 2M_3 + \frac{1}{b^2} \tilde{u}^6
\end{equation*}
whose integral curves coincide with geodesics of the $n=6$ metric (\ref{mainmet}) with mass parameter $2M_6 = \frac{1}{b^2}$ and impact parameter $\frac{1}{\tilde{b}^2} = 2M_3$. From this analysis, we get a clearer picture of what is happening. The effect of the transformation is to switch the roles of the mass (fixed) and the impact parameter (constant of integration). There do not exist two mass values $M_3$ and $M_6$ so that the totality of the geodesics from one metric will be mapped into those of the other. However, if we consider the mass term as a variable integration parameter, putting it on the same footing as $b$, then we see how the duality works.
\\In particular, the entire set of geodesics determined by the one-parameter family of metrics
\begin{equation*}
g_3 (m) = \frac{dr^2}{\left(1-\frac{2m}{r} \right)^2} + \frac{r^2}{1 - \frac{2m}{r}} d \phi^2
\end{equation*}
can be mapped into those determined by the one-parameter family
\begin{equation*}
g_6 (m) = \frac{dr^2}{\left(1-\frac{2m}{r^4} \right)^2} + \frac{r^2}{1 - \frac{2m}{r^4}} d \phi^2.
\end{equation*}
To make this idea more clear for a general $n$, we can recognize this collection of geodesics as the integral curves of a 3rd order differential equation (thus turning the mass term into a constant of integration) which can be constructed as follows:
\\From (\ref{geodeq})
\begin{equation*}
u^{1-n} u'' + u^{2-n} = nM_n
\end{equation*}
and by differentiating, we obtain
\begin{equation*}
u''' + (1-n) \frac{1}{u} (u')(u'') + (2-n) u' = 0.
\end{equation*}
Again, this equation for $n=3$ can be mapped into the $n=6$ equation via the change of coordinates
\begin{equation*}
u = \tilde{u}^{-2} \,\,\,\,\,,\,\,\,\,\, \phi = -2 \tilde{\phi}.
\end{equation*}

\textbf{Remark:} This procedure will work for any value of $n$ as long as we pick $p = -\frac{2}{n-2}$ which is the transformation implied by the Bohlin-Arnold duality.

\subsection{Zero Energy Solutions}
As said before, the zero energy solution for $n=3$ is a cardioid and for the $n=6$ case it is a Lemniscate of Bernoulli. These solutions do not get directly mapped onto each other but we can try to determine the dual curves. To obtain them, first note that the zero energy geodesic coincides with the solution of (\ref{impact}) for which $\frac{1}{b^2} \rightarrow 0$. Clearly, for any value of $n$, this gives rise to a dual curve with vanishing mass parameter which is just a projected light ray in the Minkowski case (a straight line). Thus, the zero energy solutions in the $n=3$ and $n=6$ cases with equal mass can be mapped onto each other but not directly via Bohlin-Arnold. Indeed, the zero energy curves of the equal mass black holes for any two values of $n$ can be mapped to each other in this way.

\subsection{Special Conformal Transformation}
In \cite{kothawala}, the Bohlin-Arnold duality of forces is also uncovered as a diffeomorphism of the complex plane which corresponds to a conformal transformation in real coordinates. In 2 dimensions, all real metrics are conformally flat so, from this point of view, it does not seem that this transformation is particularly special. However, by viewing it as a function on the complex plane we restrict our attention to special types of conformal transformation, which takes account of the underlying geometry.
\\Furthermore, if we wish to retain the Jacobi form of the metric, as in (\ref{jacobi}) such that the roles of the energy $\mathcal{E}$ and the potential $V(x,y)$ are switched, then we must have the transformation in the form $f(z) = z^p$ and this provides a mpa between geometries with potentials of the form $V \propto r^p$. Hence, the duality map lies in a special category of conformal transformations.

\section{Reissner-Nordstrom metrics and Duality}
One question that arises from the above work is whether a similar notion of duality exists in the case of charged black holes. To answer this, first note that the projection of a null geodesic of the metric (\ref{first}) for an arbitrary function $\Delta = \Delta(r)$ will lie in a plane, due to the inherent spherical symmetry of (\ref{first}), and will coincide with an unparametrised geodesic of the optical metric (\ref{impact}) on that plane. The differential equation describing unparametrised geodesics of (\ref{impact}) for general $\Delta = \Delta(r)$ is
\begin{equation}
(u')^2 = -u^2 \Delta + \frac{1}{b^2}
\label{general}
\end{equation}
where we have chosen the constant of integration to match up with the definition of the impact parameter from before.
\\It is clear from this equation that if we modify $\Delta$ by adding an $r^2$ term then the set of unparametrised geodesics will be unchanged. This highlights the fact that the dynamics of light rays in such static spactimes will be invariant with respect to the addition of a cosmological constant.
\\For the Reissner-Nordstrom metric in $n$ dimensions
\begin{equation*}
\Delta = 1 - \frac{2M_n}{r^{n-2}} + \frac{Q_n^2}{r^{2n-4}}
\end{equation*}
so that the equation for unparametrised geodesics of the optical metric becomes
\begin{equation}
(u')^2 + u^2 = 2M_n u^n - Q_n^2 u^{2n-2} + \frac{1}{b^2}.
\label{reissner}
\end{equation}
Now let us produce a new equation by making the change of coordinates
\begin{equation*}
u = \tilde{u}^p \,\,\,\,\,,\,\,\,\,\, \phi = p \tilde{\phi}
\end{equation*}
Then (\ref{reissner}) becomes
\begin{equation}
(\tilde{u}')^2 + \tilde{u}^2 = 2M_n \tilde{u}^{(n-2)p+2} - Q_n^2 \tilde{u}^{(2n-4)p+2} + \frac{1}{b^2} \tilde{u}^{2-2p}.
\label{nordstrom}
\end{equation}
If there is a duality as in the ST case, then we must be able to put this equation in the form of (\ref{reissner}) for some value of $n$.
\\Two of the exponents in (\ref{nordstrom}) will be equal only for $n = 0,1$ or $2$. Otherwise, we require that one of the new terms takes the place of the impact parameter i.e, one of the exponents vanishes. Since $p=1$ is trivial, this means that we have two cases to consider
\begin{equation*}
p = \frac{2}{2-n} \,\,\,\,\, \text{or} \,\,\,\,\, p = \frac{2}{4-2n}.
\end{equation*}
In the first case, (\ref{nordstrom}) becomes
\begin{equation*}
(\tilde{u}')^2 + \tilde{u}^2 = 2M_n - Q_n^2 \tilde{u}^{-2} + \frac{1}{b^2} \tilde{u}^{\frac{2n}{n-2}}
\end{equation*}
This equation resembles (\ref{reissner}) only when $n=0$ (where the duality is trivial) or when $\frac{2n}{n-2} = -6$. Hence, there is a non-trivial duality between the cases $n=-2$ and $n = \frac{3}{2}$.
\\When $p = \frac{2}{4-2n}$, (\ref{nordstrom}) becomes
\begin{equation*}
(\tilde{u}')^2 + \tilde{u}^2 = 2M_n \tilde{u} - Q_n^2 + \frac{1}{b^2} \tilde{u}^{\frac{2n-2}{n-2}}
\end{equation*}
which in the form of (\ref{reissner}) only for $n=1$ (trivial) and $n=-2$ (the duality from before).
\\In summary, we've obtained the following dual solutions:
\begin{itemize}
\item $n=0$
\\There is a duality between the $n=0$ case and the zero energy Reissner-Nordstrom solution for any value of $p$ where $M_0 + \frac{1}{2b^2}$ is the new mass parameter and $Q_n$ is the charge. Furthermore, this reduces to the ST solution for $n=1$ where $M_0 + \frac{1}{2b^2}$ becomes the mass parameter and $-Q_n^2$ is the integration constant/impact parameter.
\item $n=1$
\\Similarly, for $n=1$, we obtain a duality with the zero energy R-N solution for any value of $p$ where $M_1$ is the new mass parameter and $\sqrt{Q_n^2 - \frac{1}{b^2}}$ is the charge. This reduces to the ST for $n=-2$.
\item $n=2$
\\This is the flat case with $\Delta =$ constant and the solutions can be mapped into the zero energy solution of any ST projected null geodesic by an appropriate choice of $p$.
\item Finally, there is a duality between the cases $n = \frac{3}{2}$ and $n=-2$ where the roles of the mass, charge and impact parameters of the former are interchanged with the charge, impact parameter and mass, respectively, of the latter. However, this does not correspond to the dynamics of light rays in some optical 2-metric.
\end{itemize}
Even though none of these cases gives rise to a duality that is interesting from the optical metric point of view, we can still view the trajectories as describing particles moving in a central force of the form
\begin{equation*}
F = \frac{\alpha}{r^{n+1}} + \frac{\beta}{r^{2n-1}}
\end{equation*}
for constants $\alpha$ and $\beta$ making them still physically relevant.
\\Yet again, it is clear from the expressions (\ref{reissner}) and (\ref{nordstrom}) that the zero energy solutions can be mapped into each other for any two values of $n$ by appropriate choice of $p$.

\section{Conclusions}
We have found that the static projection of light rays of a Schwarzschild-Tangherlini metric in $n$ dimensions lie in a plane and coincide with unparametrised geodesics of an optical 2-metric. Physically, these curves also arise as non-relativistic trajectories in a central force. There is a duality due to Bohlin and Arnold between pairs of values of $n$ and we have shown that this duality implies a mapping between the totality of projected null geodesics determined by a 1-parameter family of such metrics. This notion does not extend cleanly to the charged case - the interpretation of the duality is lost in the spacetime (as $n$ is not an integer $\geq 3$) but still remains for the classical particle moving in a central force. It would be interesting to see if this notion of duality exists for other solutions of Einstein's equations. Finally, the role of the cosmological constant in spherically symmetric spacetimes of the form (\ref{general}) is illuminating. Analysis of the dynamics of light rays in such metrics does not shed light on the value of $\Lambda$ so we require other ways to measure it.

\end{document}